\documentclass[journal]{IEEEtran}
\usepackage[latin1]{inputenc}
\usepackage{graphicx}
\usepackage{amssymb,amsmath,amsfonts,amsthm}
\usepackage{float}
\usepackage{graphics}
\usepackage{xspace}
\usepackage[usenames,dvipsnames]{color}
\usepackage{amssymb, epsfig, hyperref}
\usepackage{amsmath}
\usepackage{physics}
\usepackage{epstopdf}

\usepackage[ruled,vlined,linesnumbered,noend]{algorithm2e}
\SetKwInput{KwIn}{Input}\SetKwInput{KwOut}{Output}
\usepackage[ruled,vlined,linesnumbered,noend]{algorithm2e}
\SetKwInOut{KwIn}{Input}
\SetKwInOut{KwOut}{Output}
\SetKwInOut{KwInit}{Initialization}

\begin{document}

\title{\LARGE Quantum Integrated Communication and Computing Over Multiple-Access Bosonic Channel}

\author{Ioannis Krikidis,~\IEEEmembership{Fellow, IEEE}

\thanks{I. Krikidis is with the Department of Electrical and Computer Engineering, University of Cyprus, Cyprus (e-mail: krikidis@ucy.ac.cy).}
\thanks{This work was supported by the ERC under the EU Horizon Europe programme (Grant No. 101241675, ERC PoC QUARTO).}}

\maketitle

\begin{abstract}
We investigate a quantum integrated communication and computation (QICC) scheme for a single-mode bosonic multiple-access channel (MAC) with coherent-state signalling. By exploiting the natural superposition property of the quantum MAC, a common receiver simultaneously performs over-the-air computation (OAC) on the analogue symbols transmitted by one set of devices and decodes multiple-access data from another. The joint design of the transmit power control and the receive coefficient leads to a non-convex optimization problem that maximizes computation accuracy under a prescribed sum-rate communication constraint. To address this challenge, we develop a low-complexity alternating-optimization framework that incorporates: (i) closed-form linear minimum-mean square error updates for the receive coefficient, (ii) monotonicity properties of the quantum sum-rate constraint, and (iii) projected-gradient refinements for the communication powers. The proposed QICC scheme achieves an effective computation--communication trade-off with fast convergence and low computational complexity.
\end{abstract}

\vspace{-0.25cm}
\begin{keywords}
Quantum optical communications, bosonic MAC channel, over-the-air computation, alternating-optimization.
\end{keywords}

\vspace{-0.4cm}
\section{Introduction}

\IEEEPARstart{6}{G} communication systems introduce stringent engineering requirements, including extremely high data rates and reliability, ultra-low latency, and massive connectivity \cite{TAT}. To meet these demands, both academia and industry are exploring advanced technologies and new communication paradigms that exploit the wireless medium not only for data transmission but also for additional tasks such as sensing, localization, wireless power transfer, and computation \cite{LIM}. In this context, the concept of over-the-air computation (OAC) leverages the natural superposition property of the wireless multiple-access channel (MAC) to compute specific classes of functions (nomographic functions) directly ``in the air,'' without requiring individual signal recovery at the receiver \cite{CAO,LIU}. OAC has been extensively investigated in the literature under a wide range of system configurations and design perspectives. A more recent and enhanced framework, termed integrated communication and computing (ICC), enables simultaneous analog computation and digital data decoding at a common receiver. In ICC, the receiver computes a function of the analog signals transmitted by one set of devices while concurrently decoding digital messages from another set \cite{YE}. To effectively mitigate the multi-user interference arising between the computation and communication tasks, advanced interference-management techniques, including dirty-paper coding, have been incorporated, leading to significant performance improvements \cite{ROC}.

On the other hand, quantum communications exploit the non-classical properties of quantum mechanics to achieve fundamentally enhanced security, higher information capacity, and performance benefits that are unattainable with conventional classical systems \cite{DJO}. Here, we consider the case where classical information is embedded in quantum states and transmitted through quantum channels. In this context, the natural extension of the classical MAC is the quantum MAC, whose fundamental limits have been characterized in \cite{WIN} for the general setting with arbitrary receiver measurements, and in \cite{YEN} for the special case of bosonic channels. Recently, there has been increasing interest in importing communication paradigms originally designed for electromagnetic wireless systems into the quantum domain. For instance, \cite{KRI} introduces wireless-powered quantum communications, while \cite{CON} studies integrated quantum sensing and communication.  

In this letter, we extend the ICC concept to the quantum domain and introduce a new communication paradigm, termed quantum integrated communication and computation (QICC). We consider a bosonic MAC where a common receiver performs over-the-air computation on the analogue signals of the computation devices and, in parallel, decodes the classical messages conveyed by the communication devices. Both tasks are enabled by encoding all signals into quantum coherent states, which are naturally superimposed by the bosonic MAC. The joint design is formulated as a non-convex optimization problem that minimizes the mean-square error (MSE) of the computation task under a sum-rate constraint (expressed in terms of the von Neumann entropy) with respect to the transmit powers and the receive coefficient. The proposed solution exploits the problem structure through an alternating-optimization (AO) framework that iteratively performs: (i) closed-form linear minimum MSE (LMMSE) updates for the receive coefficient, (ii) one-dimensional bisection for the computation powers, and (iii) projected-gradient updates for the communication powers, while ensuring low complexity and convergence guarantees. Numerical results reveal a fundamental communication--computation trade-off and confirm the efficiency of the proposed framework. The objective of this work is not to compare with classical ICC systems, which operate over fundamentally different channels, but to introduce the concept of ICC in bosonic quantum systems using a coherent-state signaling framework with a simple measurement baseline.

\vspace{-0.4cm}
\section{System model \& problem formulation}

We consider a bosonic single-mode optical uplink system that jointly supports OAC and MAC communication. The system consists of $K$ OAC devices, $M$ communication devices, and a common receiver that performs quantum measurements to obtain the observations used for OAC estimation and data decoding. All transmitters employ coherent-state signalling, where each device prepares a coherent state by applying a displacement operation to the vacuum \cite{DJO}. The OAC devices embed their analogue computation symbols $s_k$, $k=1,\ldots,K$, into coherent states, while the communication devices encode their digital symbols $d_m$, $m=1,\ldots,M$, in the same manner. Due to the linearity of the bosonic channel, all transmitted coherent states superpose at the receiver input through the coherent-state MAC \cite[Sec.~II.B]{YEN}.

For simplicity, we assume $s_k \in \mathbb{C}$ and $d_m \in \mathbb{C}$ are zero-mean, unit-power complex random variables {\it i.e.,} $\mathbb{E}[s_k]=0$ and $\mathbb{E}[|s_k|^2] = 1$, $\mathbb{E}[d_m]=0$ and $\mathbb{E}[|d_m|^2]=1$; the symbols are mutually uncorrelated {\it i.e.},  $\mathbb{E}[s_k s_i^{\mathrm{H}}] = 0$ $\forall k\neq i$, and $\mathbb{E}[d_m d_i^{\mathrm{H}}]=0$  $\forall m\neq i$. Let $\ket{\sqrt{g_k} s_k}$, $\ket{\sqrt{P_m}\, d_m}$ denote the transmitted coherent states of the $k$-th computation device and $m$-th communication device, respectively, where $0\le g_k\le P_c$ and $0\le P_m\leq P_t$ denote their associated power control coefficients. All transmitted coherent states interfere through the bosonic MAC \cite{YE} with device-specific transmissivities $\eta_i \in (0,1]$ satisfying 
$\sum_{i=1}^{K+M} \eta_i = 1$, yielding a single output mode in the pure 
coherent state
\vspace{-0.2cm}
\begin{align}
\ket{Y_{\mathrm{out}}}=
\ket{\sum_{k=1}^K \sqrt{\eta_k g_k}\, s_k 
	+ \sum_{m=1}^M \sqrt{\eta_{K+m}P_m}\, d_m}.
\end{align}
The receiver performs heterodyne detection to obtain the classical observations used for the computation task, which results in the equivalent classical input-output channel model
\vspace{-0.2cm}
\begin{align}
y = \sum_{k=1}^K \sqrt{\eta_k g_k}\, s_k
+ \sum_{m=1}^M \sqrt{\eta_{K+m}P_m}\, d_m
+ z,
\label{MAC1}
\end{align}
where $z \sim \mathcal{CN}(0,N_0)$ models thermal/quantum measurement noise. By exploiting this inherent aggregation of the optical field in the output mode, the receiver simultaneously (i) employs heterodyne detection to obtain the classical observations for estimating the desired OAC function ({\it i.e.,} $S=\sum_{k=1}^{K}s_k$), and (ii) supports reliable decoding of the communication symbols under the bosonic MAC sum-rate bound.

The receiver adopts a linear estimator $\hat{S}=h y$ with $h\in\mathbb{C}$, and OAC computation accuracy can be expressed in terms of MSE. Concurrently, the communication users operate over a bosonic MAC, for which the achievable rate region is characterized by the capacity results for bosonic MAC~\cite[Eq.~12]{YEN}. In this context, the sum-rate expression corresponds to the quantum information-theoretic limit of the bosonic MAC and therefore serves as a fundamental upper bound on the achievable communication performance, independent of the specific receiver implementation. For simplicity, we require that the communication devices collectively achieve at least a prescribed minimum sum-rate. Therefore, we are interested in jointly designing the OAC and communication power control coefficients $(\{g_k\}_{k=1}^{K}, \{P_m\}_{m=1}^{M})$ and the receive coefficient $h$ in order to minimize the computation distortion while guaranteeing a reliable sum-rate. The resulting optimization problem can be written as
\begin{align}
{\bf [P1]}\;		&\min_{\{g_k\},\, \{P_m\},\, h} \quad 
		 \mathsf{MSE}\!\left(\{g_k\}, \{P_m\}, h\right) \nonumber\\
		 \quad  &\text{s.t.}\quad R^{\mathrm{sum}}
		\;\le\; g\!\left( N_{\mathrm{sig}} + N_{\mathrm{eff}} \right)
		- g\!\left( N_{\mathrm{eff}} \right), 
		\label{opt-cap-const} \\[0.4em]
		& 0\le g_k \le P_c,
		\qquad k = 1,\ldots,K,
		\label{power1} \\[0.4em]
		& 0\le P_m \le P_t, 
		\qquad m = 1,\ldots,M,
		\label{power2}
\end{align}
with
\begin{align}
&\mathsf{MSE}(\{g_k\},\{P_m\}, h)
=\mathbb{E}[|S-\hat{S}|^2] \nonumber \\
&\;\;=\sum_{k=1}^{K} \big| h\sqrt{\eta_kg_k} - 1 \big|^2+ |h|^2 \left( N_{\mathrm{sig}} + N_0 \right), \label{mse1}
\end{align}
where $R^{\mathrm{sum}}$ denotes the requested sum-rate and thus \eqref{opt-cap-const} represents the communication constraint \cite[Eq. 12]{YEN}; \eqref{power1} and \eqref{power2} refer to the power constraints for the computation and communication devices, respectively; $N_{\mathrm{sig}} = \sum_{m=1}^{M} \eta_{K+m} P_m$ denotes the total received signal power from the communication devices, $N_{\mathrm{eff}} = N_{0} + \sum_{k=1}^{K} \eta_k g_k$ is the effective noise-plus-interference term seen by the communication decoding; $g(x) = (x+1)\log_2(x+1) - x\log_2(x)$ denotes the von Neumann entropy (in bits per channel use) of a single-mode thermal state with mean photon number $x$; the achievable rate in \eqref{opt-cap-const} follows from the classical capacity of phase-insensitive bosonic Gaussian channels, $C = g(N_{\mathrm{sig}}+N_{\mathrm{eff}}) - g(N_{\mathrm{eff}})$ under an energy constraint \cite[Eq. 1]{HOL2}

\vspace{-0.3cm}
\section{Quantum integrated communication and computing design} 

The joint design problem in [P1] corresponds to a QICC system and results in a non-convex optimization formulation. To handle this intractability, we adopt an AO framework in which each block of variables is optimized while keeping the remaining ones fixed. This approach exploits the problem structure and enables an efficient iterative solution that converges to a stationary point.
\vspace{-0.4cm}
\subsection{Optimization of the linear estimator scalar $h$}\label{sec_h}

Given the OAC and communication power control coefficient $\{g_k\}$ and $\{P_m\}$, 
the optimal LMMSE receive coefficient $h$ minimizes 
$\mathsf{MSE} = \mathbb{E}\{|S - h y|^2\}$ and is given by
\begin{align}
h^\star = \frac{\mathbb{E}\{S y^\ast\}}{\mathbb{E}\{|y|^2\}}\;=
\frac{\sum_{k=1}^K \sqrt{\eta_k g_k}}
{\sum_{k=1}^K \eta_k g_k 
	+ N_{\mathrm{sig}}
	+ N_0 }. \label{h_opt}
\end{align}
with $\mathbb{E}\{S y^\ast\} = \sum_{k=1}^K \sqrt{\eta_k g_k}$ and $\mathbb{E}\{|y|^2\} 
= \sum_{k=1}^K \eta_k g_k 
+ N_{\mathrm{sig}} + N_0$.
By substituting $h^\star$ back into the MSE expression in \eqref{mse1}, we obtain 
\vspace{-0.2cm}
\begin{align}
	&\mathsf{MSE}(\{g_k \},\{P_m\})= K -
	\frac{\left(\sum_{k=1}^{K} \sqrt{\eta_k g_k}\right)^{2}}
	{
		\sum_{k=1}^{K} \eta_k g_k 
		+ N_{\mathrm{sig}} 
		+ N_0 } .
	\label{eq:MSE_min}
\end{align}

\noindent{\it Minimum/maximum MSE:} From the above expression, the minimum MSE is attained when the communication devices remain inactive ({\it i.e.,} when $N_{\mathrm{sig}} = 0$). In this case, the MSE objective depends solely on the computation devices, which optimally transmit at their maximum power (see Appendix \ref{apA}). Consequently, the minimum achievable MSE simplifies to $\mathsf{MSE}_{\mathrm{min}}=K-\frac{\left(\sum_{k=1}^K \sqrt{\eta_k P_c} \right)^2}{\sum_{k=1}^K\eta_kP_c+N_0}$. On the other hand, the maximum MSE is achieved when $g_k=0\;\forall k$ and $\mathsf{MSE}_{\mathrm{max}}=K$.

\vspace{-0.3cm}
\subsection{Optimization of the communication power $\{P_m\}$}

Given the OAC power control coefficients $g_k$ and the LMMSE coefficient  $h^\star$, we compute the communication power coefficients $\{P_m\}$. However, the MSE expression depends on $\{P_m\}$ only through the aggregate received communication power $N_{\mathrm{sig}}$ which enters the denominator of the reduced MSE in~\eqref{eq:MSE_min}; therefore, the individual power variables $\{P_m\}$ do not affect the MSE beyond their contribution to $N_{\mathrm{sig}}$. At the same time, the communication task imposes the sum-rate constraint in \eqref{opt-cap-const}; since $g(\cdot)$ is strictly increasing, the minimal communication power that satisfies~\eqref{opt-cap-const} is given by
\begin{equation}
g\!\left(N_{\mathrm{sig}}^{\star} + N_{\mathrm{eff}}\right)
- g\!\left(N_{\mathrm{eff}}\right)
=
R^{\mathrm{sum}},
\label{eq:Nsig_equation}
\end{equation}
subject to the power budget
\vspace{-0.2cm}
\begin{equation}
	0 \le N_{\mathrm{sig}}^{\star}
	\le  N_{\mathrm{sig}}^{\max}=\sum_{m=1}^{M} \eta_{K+m} P_t.
	\label{eq:Nsig_range}
\end{equation}
Because the left-hand side of~\eqref{eq:Nsig_equation} is a strictly 
increasing function of $N_{\mathrm{sig}}$, the value $N_{\mathrm{sig}}^\star$
can be efficiently obtained via a {\it one-dimensional bisection search}.
It is worth noting that any feasible set of communication powers $\{P_m\}$ that satisfies
$N_{\mathrm{sig}}^\star=\sum_{m=1}^{M} \eta_{K+m} P_m$ is equivalent from the perspective of the MSE. This step eliminates the communication powers from~[P1] and leaves a reduced
optimization problem over the OAC powers $\{g_k\}$ only.

\noindent{\it Feasibility:} Since $N_{\mathrm{eff}} \geq N_0$ and the von Neumann entropy 
$g(x)$ is a monotonically increasing function of its argument, 
the maximum sum rate that the system can support is $R_{\mathrm{max}}^{\mathrm{sum}}= g\!\left(\sum_{m=1}^{M} \eta_{K+m} P_t + N_0\right)- g(N_0)$.

\vspace{-0.4cm}
\subsection{Optimization of the computation power $\{g_k\}$}\label{secc}

By substituting $h^\star$ from~\eqref{h_opt} and $N_{\mathrm{sig}}^\star$
from~\eqref{eq:Nsig_equation} into the MSE expression in~\eqref{mse1}, we
obtain the reduced optimization problem
\begin{align}
	{\bf [P2]}\quad	
	\min_{\{g_k\}} \;\;&
	\mathsf{MSE}\big(\{g_k\}\big)
	\label{eq:opt_g}\\[2pt]
	\text{s.t.}\quad 
	& R^{\mathrm{sum}}
	\;\le\; g\!\left( N_{\mathrm{sig}}^{\max} + N_{\mathrm{eff}} \right)
	- g\!\left( N_{\mathrm{eff}} \right), 
	\label{c1}\\
	& 0 \le g_k \le P_c,\quad k=1,\ldots,K.
\end{align}
where the constraint in \eqref{c1} corresponds to the best-case communication power; since $g(x)$ is strictly increasing for $x>0$, the feasibility condition follows from the maximum received communication power. The formulation in [P2] is non-convex due to the fractional quadratic form of the MSE objective and the non-linear dependence of $N_{\mathrm{eff}}$ on $\{g_k\}$. To solve [P2], we adopt a projected-gradient method (where a gradient descent step on the MSE objective is followed by a projection onto the feasible power region \cite[Ch. 2]{BER}) within the AO framework. At each iteration $n$, the OAC powers $\{g_k\}$ are updated via a projected-gradient step. We first take an unconstrained gradient descent step
\begin{align}
	\tilde{g}_k = g_k^{(n)} - \mu\,\frac{\partial \mathrm{MSE}}{\partial g_k},
	\label{eq:pg_update}
\end{align}
where $\mu>0$ is a suitable stepsize, and clip each component to the
individual power budget $\bar{g}_k = \min\{P_c,\max\{0,\tilde{g}_k\}\}$.

If the gradient step satisfies the sum-rate feasibility constraint $\sum_k \eta_k \bar{g}_k \le \Gamma_{\max}$ we have $g_k^{(n+1)} = \bar g_k$, otherwise we project the point back onto the corresponding half-space by applying Euclidean projection (onto the linear constraint $\sum_k \eta_k g_k \le \Gamma_{\max}$, which admits a closed-form solution) {\it i.e.,} $g_k^{(n+1)} = \bar g_k - \eta_k\, \frac{\sum_j \eta_j \bar g_j - \Gamma_{\max}} {\sum_j \eta_j^2}$, where  $\Gamma_{\max} = N_{\mathrm{eff}}^{\max} - N_0$ denote the maximum allowable aggregated OAC contribution imposed by the sum-rate constraint; $\Gamma_{\max}$ is obtained numerically by solving
the one-dimensional equation (via bisection) $g\!\left( N_{\mathrm{sig}}^{\max} + N_{\mathrm{eff}}^{\max} \right)-g\!\left( N_{\mathrm{eff}}^{\max} \right)= R^{\mathrm{sum}}$.
This guarantees that all iterates satisfy $0\le g_k \le P_c$ and the
feasibility condition in~\eqref{c1}.

\vspace{-0.3cm}
\subsection{Overall Alternating-Optimization Framework}

The updates derived in the previous subsections naturally lead to an 
AO procedure in which the variable blocks 
$\{g_k\}$, $N_{\mathrm{sig}}$, and $h$ are updated sequentially. 
At each iteration $n$, (i) the LMMSE estimator $h$ is obtained in closed 
form using~\eqref{h_opt}, (ii) the aggregate communication power 
$N_{\mathrm{sig}}^\star$ is computed via the one-dimensional bisection 
solution of~\eqref{eq:Nsig_equation}, and (iii) the OAC powers $\{g_k\}$ 
are refined through the projected-gradient update in~\eqref{eq:pg_update}.  Each block update is optimal for its corresponding subproblem and yields a monotonic decrease of the overall MSE objective in~[P1]. Since the MSE objective is non-negative, and each block update produces a continuous and monotonic decrease of the objective over a compact feasible set, the AO procedure is guaranteed to converge to a stationary point \cite{RAZ}.

\noindent{\it Complexity:} The update of $h$ has closed-form complexity $\mathcal{O}(K)$, the computation of $N_{\mathrm{sig}}^\star$ requires a one-dimensional bisection search of 
complexity $\mathcal{O}(\log(1/\varepsilon_{\mathrm{MSE}}))$ \cite{BUR}, and the projected 
gradient step for the $K$ powers has complexity $\mathcal{O}(K)$. 
Therefore, each AO iteration has overall complexity $\mathcal{O}(K + \log\!\frac{1}{\varepsilon_{\mathrm{MSE}}})\approx \mathcal{O}(K)$, which makes the proposed method lightweight and well-suited for practical 
implementations. To summarize the overall approach, Algorithm~\ref{alg:AO_PG} provides the 
full pseudocode of the proposed iterative method.

\begin{algorithm}[t]
	\small
	\caption{Alternating Optimization for QICC}
	\label{alg:AO_PG}
	\KwIn{$R^{\mathrm{sum}}$, $\mu$, $P_c$, $P_t$, tolerances $\varepsilon_{\mathrm{AO}}$, $\varepsilon_{\mathsf{MSE}}$, maximum iterations $N_{\mathrm{max}}$.}
	\KwOut{OAC powers $\{g_k\}$, receive coefficient $h$, aggregate communication power $N_{\mathrm{sig}}$.}
	
	\KwInit{Choose feasible OAC powers $\{g_k^{(0)}\}_{k=1}^K$; set $n \leftarrow 0$; compute $\mathsf{MSE}^{(0)}$. Pre-compute $\Gamma_{\max}$ from the one-dimensional equation in Sec.~\ref{secc} (via bisection).}
	
	\Repeat{$\big|\mathsf{MSE}^{(n)} - \mathsf{MSE}^{(n-1)}\big| \le \varepsilon_{\mathrm{AO}}$ \rm{or} $n = N_{\max}$}{
		
		\tcp{Update linear estimator}
		Compute $h^{\star (n)}$ using the closed-form expression in~\eqref{h_opt} \;
		
		\tcp{Effective noise for MAC decoder}
		$N_{\mathrm{eff}}^{(n)} \leftarrow N_{0} + \sum_{k=1}^{K} \eta_k g_k^{(n)}$ \;
		
		\tcp{Aggregate communication power from sum-rate constraint}
		Compute $N_{\mathrm{sig}}^{\star (n)}$ as the solution of
		\[
		g\big(N_{\mathrm{sig}}^{\star} + N_{\mathrm{eff}}^{(n)}\big)
		- g\big(N_{\mathrm{eff}}^{(n)}\big)
		= R^{\mathrm{sum}}
		\]
		subject to
		$
		0 \le N_{\mathrm{sig}}^{\star}
		\le \sum_{m=1}^{M} \eta_{K+m} P_t,
		$
		via 1-D bisection with tolerance $\varepsilon_{\mathrm{MSE}}$ \;
		
		\tcp{Projected-gradient update of OAC powers}
		Compute $\partial\mathrm{MSE}\big(\{g_k^{(n)}\}\big)/\partial g_k $ (cf. Appendix \ref{apB}) \;
		
		\For{$k = 1,\dots,K$}{
			\tcp{Unconstrained gradient step}
			$\tilde g_k^{(n+1)} \leftarrow g_k^{(n)}
			- \mu \, \partial \mathrm{MSE}\big(\{g_k^{(n)}\}\big)/\partial g_k$ \;
			
			\tcp{Clip to individual box constraint}
			$\bar g_k^{(n+1)} \leftarrow \min\!\big\{P_c,\; \max\{0,\tilde g_k^{(n+1)}\}\big\}$ \;
		}
		
		\tcp{Check aggregate OAC constraint and project if needed}
		$\Gamma^{(n+1)} \leftarrow \sum_{k=1}^K \eta_k \bar g_k^{(n+1)}$ \;
		
		\eIf{$\Gamma^{(n+1)} \le \Gamma_{\max}$}{
			\tcp{Feasible: accept clipped powers}
			$g_k^{(n+1)} \leftarrow \bar g_k^{(n+1)}$ for all $k$ \;
		}{
			\tcp{Infeasible: Euclidean projection onto $\sum_k \eta_k g_k \le \Gamma_{\max}$}
			$\displaystyle \Delta \leftarrow \frac{\sum_{j=1}^K \eta_j \bar g_j^{(n+1)} - \Gamma_{\max}}
			{\sum_{j=1}^K \eta_j^2}$ \;
			$g_k^{(n+1)} \leftarrow \bar g_k^{(n+1)} - \eta_k \Delta$, \quad $k=1,\ldots,K$ \;
		}
		
		\tcp{Update objective and iteration index}
		$\mathsf{MSE}^{(n+1)} \leftarrow \mathsf{MSE}\big(\{g_k^{(n+1)}\}\big)$ \;
		$n \leftarrow n+1$ \;
	}
\end{algorithm}

\vspace{-0.2cm}
\section{Numerical studies}

Computer simulations are carried out to evaluate the performance of the proposed QICC scheme. The simulation setup assumes $\mu=10^{-3}$, $\varepsilon_{\mathrm{AO}}=\varepsilon_{\mathsf{MSE}}=10^{-6}$, $N_{\mathrm{max}}=1000$, $N_0=2$, $\eta_k=0.6/K$ for $k=1,\ldots,K$, $\eta_{K+m}=0.4/M$ with $m=1,\ldots,M$, and $P_c=P_t$.

\begin{figure}
\centering
	\includegraphics[width=0.74\linewidth]{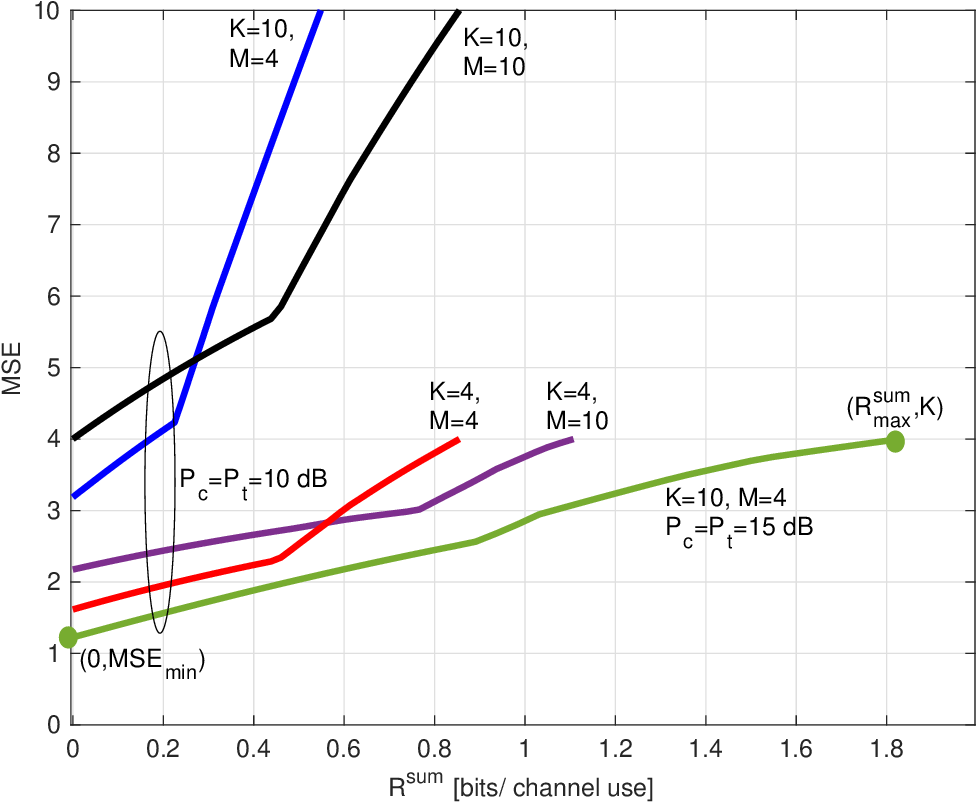}
		\vspace{-0.35cm}
\caption{Achievable MSE--sum-rate performance for different $(K,M)$ system configurations; the circle markers denote the boundary points $(0,\mathsf{MSE}_{\mathrm{min}})$ and $(R^{\mathrm{sum}}_{\mathrm{max}},K)$.
  }
	\label{fig1}
\end{figure}

Fig.~\ref{fig1} illustrates the achievable MSE--sum-rate performance for different $(K,M)$ configurations and power budgets. As expected, the MSE attains its minimum when no communication rate is required and increases monotonically with the sum-rate constraint, reflecting the fundamental computation--communication tradeoff. We further observe that curves with the same number of computation devices $K$ but different numbers of communication devices $M$ may intersect. For small sum-rate constraints, the configuration with fewer communication devices achieves slightly lower MSE, as the design operates in a computation-dominated regime. In contrast, for larger sum-rate requirements, the setup with larger $M$ becomes advantageous, since distributing the communication load across more transmitters allows the system to satisfy stringent rate constraints with a smaller degradation in computation accuracy. Finally, increasing the transmit powers $(P_c,P_t)$ enables operation at higher sum rates while simultaneously achieving lower MSE levels. Note that the reported MSE values correspond to the absolute estimation error and naturally scale with the number of computation devices, while the coexistence of communication and computation signals inherently introduces mutual interference.

\begin{figure}
	\centering
	\includegraphics[width=0.74\linewidth]{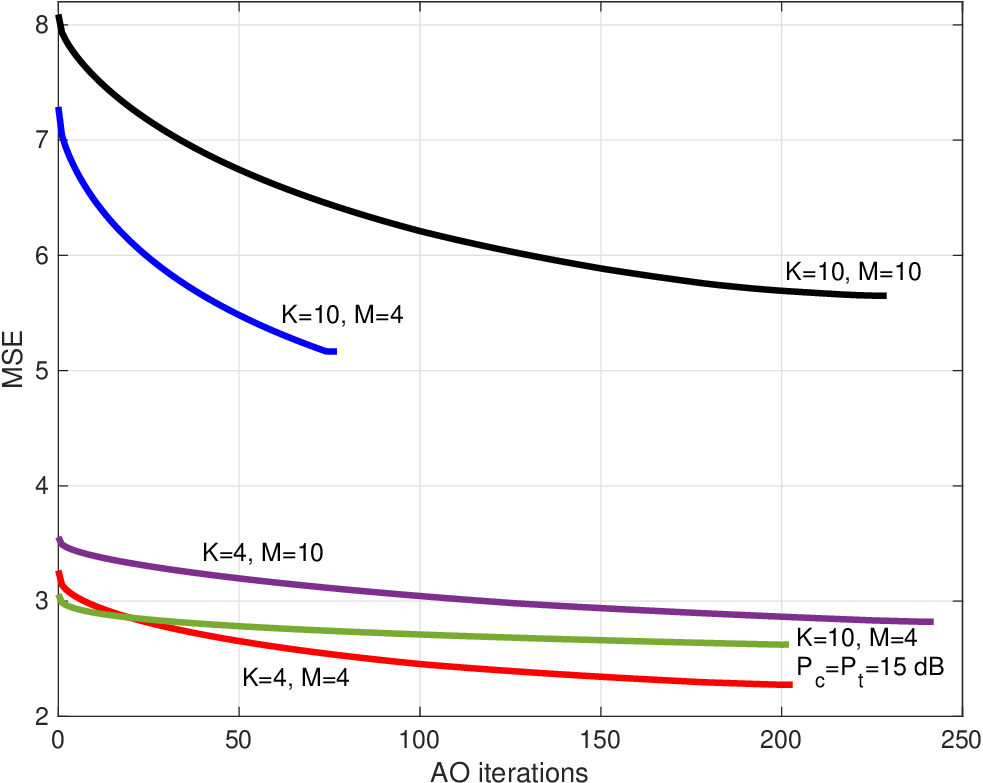}
	\vspace{-0.35cm}
\caption{Convergence behavior of the proposed AO algorithm for various 
	$(K,M)$ configurations and
	$R_{\mathrm{sum}} = R_{\mathrm{sum}}^{\max}/2$.}
	\label{fig2}
\end{figure}

Fig.~\ref{fig2} illustrates the convergence trajectory of the proposed AO framework for the same configurations examined in Fig.~\ref{fig1}. In all cases, the MSE exhibits a smooth and monotonic decrease and converges well before reaching the maximum number of allowed iterations. This confirms that the AO scheme is numerically stable, has low computational complexity, and is well suited for real-time QICC implementations. 

Overall, QICC introduces a new quantum paradigm that unifies communication and computation in the quantum optical domain. Our results reveal a trade-off between computation accuracy and communication throughput, while the proposed algorithm achieves this balance with low complexity and fast convergence. Future extensions may incorporate non-classical quantum resources, such as entanglement or squeezing, as well as alternative quantum MAC configurations to explore potential performance gains, while also considering practical hardware impairments and nonlinear effects in more realistic system scenarios.

\appendices

\vspace{-0.45cm}
\section{Minimum MSE for the special case with $N_{\mathrm{sig}}=0$}\label{apA}

Let $x_k \triangleq \sqrt{\eta_k g_k}$ so that $0 \le x_k \le \sqrt{\eta_k P_c}$.
To verify that no interior point can be optimal, we examine the objective
along the ray obtained by scaling all $x_k$ by a common factor $t>1$.
For any feasible $\{x_k\}$ and any $t>1$ such that 
$t x_k \le \sqrt{\eta_k P_c}$ for all $k$, consider the one-dimensional
function $f(t)=\frac{\big(\sum_k t x_k\big)^2}{\sum_k t^2 x_k^2 + N_0}$. A direct calculation shows that $f'(t)>0$ for all $t>0$, and thus $f(t)$
is strictly increasing for all feasible $t>1$.  Therefore the objective
is maximized by taking each $x_k$ at its largest feasible value,
namely $x_k^\star=\sqrt{\eta_k P_c}$, which implies $g_k^\star=P_c$.
Substituting these values into the reduced MSE expression yields the
closed-form minimum MSE stated in Section~\ref{sec_h}.

\vspace{-0.45cm}
\section{Computation of the MSE gradient}\label{apB}
Let $A \triangleq \sum_{k=1}^K \sqrt{\eta_k g_k}$ and 
$D \triangleq \sum_{k=1}^K \eta_k g_k + N_{\mathrm{sig}}^\star + N_0$, 
so that the reduced MSE in \eqref{eq:MSE_min} becomes $\mathrm{MSE}(\{g_k\})
= K - \frac{A^{2}}{D}$ For real $g_k \ge 0$, the corresponding gradient is
\begin{equation}
	\frac{\partial\,\mathsf{MSE}}{\partial g_k}
	=
	-\,\frac{
		D\,A\,\frac{\sqrt{\eta_k}}{\sqrt{g_k}}
		- A^{2}\eta_k
	}{D^{2}},
	\qquad k=1,\ldots,K,
\end{equation}
and this expression is used in the projected-gradient update of the 
OAC powers in~\eqref{eq:pg_update}.

\end{document}